\documentclass[journal]{vgtc}                     % final (journal style)
\onlineid{1230}

\vgtccategory{Research}

\vgtcpapertype{theory/model}

\author{%
  \authororcid{Xi Duan}{0009-0003-4926-6547},
  \authororcid{Yiwei Lin}{0009-0004-5807-9492},
  \authororcid{Shiqing Xin}{0000-0001-8452-8723},
  \authororcid{Aoying Wang}{0009-0004-9541-5774},
  \authororcid{Yucheng Wang}{0000-0002-8336-1519},
  \authororcid{Changhe Tu}{0000-0002-1231-339},
  and
  \authororcid{Qiong Zeng}{0000-0002-2827-8261}
}

\authorfooter{
  \item
  Xi Duan, Yiwei Lin, Shiqing Xin, and Changhe Tu are with Shandong University. E-mail: duanximail@gmail.com; ywlin1499@gmail.com; xinshiqing@sdu.edu.cn; changhe.tu@gmail.com.
  
  \item
  Aoying Wang is with Ocean University of China. E-mail: wangaoying@stu.ouc.edu.cn.
  
  \item
  Yucheng Wang is with Laoshan Laboratory. E-mail: ycwang@lsnl.cn.

  \item Qiong Zeng, the corresponding author, is with Shandong University and Laoshan Laboratory. Email: qiong.zn@sdu.edu.cn.
}

\title{Beyond the Good, the Bad, and the Ugly: \\Colormap Assessment through Data-Aware Perceptual Metric}

\abstract{%
  Continuous colormaps are widely used to visualize scalar fields, and their quality is typically evaluated using measures such as discriminative power and uniformity. Existing measures primarily characterize the intrinsic perceptual properties of the colormap itself, largely independent of the underlying data distribution.
  In practice, however, user perception arises not from the colormap alone but from the visualization generated by mapping data values through the colormap. The perceptual differences that users actually experience depend jointly on the colormap and the underlying data. We propose a data-aware formulation that complements existing data-independent colormap assessment approaches. 
  Rather than analyzing the colormap in isolation, we model the color-encoded visualization as a composite mapping from the spatial domain of the data to perceptual color space. This approach yields data-aware counterparts of established measures, including discriminative power, uniformity, and smoothness, as well as additional measures such as perceptual anisotropy and degeneracy. 
  We validate the proposed measures against both the existing data-independent framework and empirical results from perceptual studies, extend the formulation to 2D colormaps, and demonstrate its integration into colormap optimization. An interactive system is provided for exploring colormap assessment under varying data distributions. Results demonstrate that our formulation offers a principled foundation for data-aware colormap assessment and design.
}


\teaser{
  \centering
  \includegraphics[width=\linewidth]{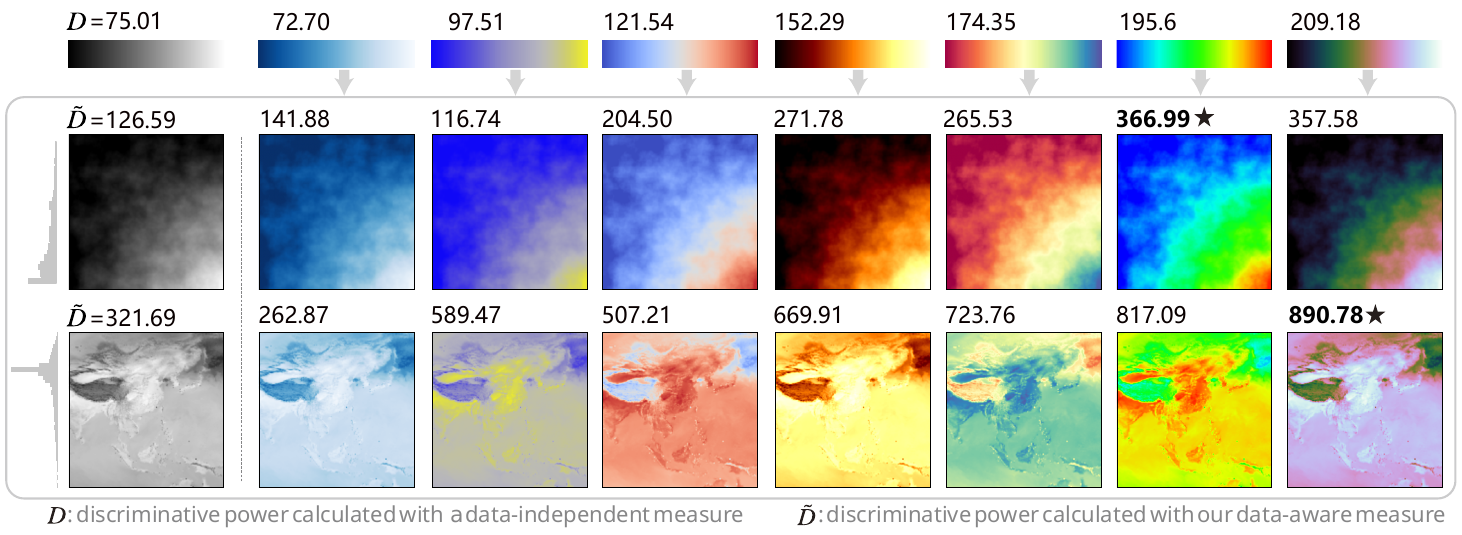}
  \caption{\textbf{Colormaps and corresponding visualizations ranked by discriminative power.} \textbf{Top:} data-independent ranking using the measure of Bujack~\etal~\cite{Bujack18}. \textbf{Middle:} data-aware ranking on a synthetic Perlin noise dataset (D1). \textbf{Bottom:} data-aware ranking on a real-world temperature dataset (D7). Both datasets exhibit skewed value distributions (histograms shown at left). The data-aware rankings differ substantially from the data-independent baseline, as most data values concentrate in a narrow range where different colormaps exhibit different perceptual properties.}
  \label{fig:teaser}
}

\graphicspath{{figs/}{figures/}{pictures/}{images/}{./}} % where to search for the images
\usepackage{tabu}                      % only used for the table example
\usepackage{booktabs}                  % only used for the table example
\usepackage{lipsum}                    % used to generate placeholder text
\usepackage{mwe}                       % used to generate placeholder figures
\usepackage[pagebackref,bookmarks]{hyperref}
\usepackage{mathptmx}                  % use matching math font
\usepackage{textcomp}                  % use better special symbols
\usepackage{times}                     % we use Times as the main font
\usepackage{cite}                      % needed to automatically sort the references
\usepackage{mathtools}
\usepackage{caption}
\usepackage{color}
\usepackage{enumitem}
\usepackage{amsmath}
\usepackage{amsfonts}
\usepackage{amssymb}
\usepackage{graphics}
\usepackage[utf8]{inputenc}
\usepackage{amsmath,bm}
\usepackage{multirow}
\usepackage{multicol}
\usepackage{subcaption}
\usepackage{tcolorbox}
\usepackage{fixdif, derivative}
\usepackage{algorithm}
\usepackage{algpseudocode}
\tcbuselibrary{breakable}
\newif\ifnotes
\notesfalse % hide notes
\newif\ifedited
 \editedtrue % display edits
\editedfalse % hide notes
\definecolor{editcolor}{rgb}{0, 0, 0}
\definecolor{viseditcolor}{rgb}{1, 0, 0}
\definecolor{black}{rgb}{0,0,0}

\def\etal{{\textit{et~al.}}}

\usepackage{soul}

\usepackage{afterpage}
\usepackage{xspace}
\usepackage[nameinlink,capitalise]{cleveref}
  \crefname{table}{Tab.}{Tabs.}
  \Crefname{table}{Table}{Tables}
  \crefname{section}{Sec.}{Secs.}
  \Crefname{Section}{Section}{Sections}
  \crefname{figure}{Fig.}{Figs.}
  \Crefname{Figure}{Fig.}{Figs.}
  \Crefname{equation}{Eq.}{Eqs.}

\usepackage{xcolor}
\usepackage{tikz}
\usepackage{caption}

\DeclareRobustCommand{\legendline}[1]{%
  \tikz[baseline=-0.6ex]\draw[#1, line width=1pt] (0,0) -- (0.3cm,0);%
}

\begin{document}
%%%%%%%%%%%%%%%%%%%%%%%%%%%%%%%%%%%%%%%%%%%%%%%%%%%%%%%%%%%%%%%%
%%%%%%%%%%%%%%%%%%%%%% START OF THE PAPER %%%%%%%%%%%%%%%%%%%%%%
%%%%%%%%%%%%%%%%%%%%%%%%%%%%%%%%%%%%%%%%%%%%%%%%%%%%%%%%%%%%%%%%

\maketitle

\section{Introduction}
\label{sec:Introduction}
Continuous colormaps are used to visualize scalar data by mapping numerical values to colors in a perceptual color space~\cite{ware2021}. Effective colormap design is essential for enabling users to accurately perceive values, gradients, and anomalies in scalar fields across a wide range of scientific and analytical applications~\cite{ParaView05}. However, designing effective colormaps is complex due to the relationships among human perception, colormap design, analytical tasks, and the characteristics of the underlying data~\cite{schulzewollgast2005, Tominski08, Reda18, Ware19, Zeng22}. Numerous empirical studies have shown that poorly designed colormaps can distort perceived gradients, introduce visual artifacts, and affect users' performance in visualization tasks~\cite{Rogowitz98, Dasgupta20, Crameri20}. Assessing the perceptual quality of colormaps has therefore become a central concern in visualization research.

Over the years, a wide range of design rules have been proposed for what constitutes a good colormap, including monotonicity, continuity, linearity, constant speed, and low curvature, among others~\cite{Levkowitz92, Ware88, Nardini21b, Moreland09}. However, the meaning of many of these terms has been ambiguous or open to interpretation, with different authors sometimes using the same term to describe different concepts. Bujack~\etal~\cite{Bujack18} addressed this fragmentation by proposing a specified taxonomy and providing unambiguous mathematical definitions for the most common design rules, grounded in mutual distances and speed between colors in perceptual color space. The influential work established a shared language and a principled foundation for colormap evaluation that has since been widely adopted. By design, these definitions are independent of the colorspace, task, or data under consideration, ensuring generality across applications.

However, data independence also means that these measures do not reflect how a colormap performs when applied to a specific dataset. In practice, the perceptual differences that users experience depend jointly on both the colormap and the underlying data. 
\cref{fig:teaser} illustrates this with two datasets: a synthetic Perlin noise field (D1) and a real-world ocean temperature field (D7), each with a distinct value distribution. The same colormap produces noticeably different perceptual outcomes on the two datasets, and different colormaps applied to the same dataset reveal different levels of detail in the underlying structure. A similar perspective appears in the work of Bujack~\etal~\cite{Bujack26}, who assess visualization quality by how accurately the original data can be reconstructed from the visualization; their colormapping examples further motivate evaluating colormaps in conjunction with data rather than in isolation.
The interplay between data and colormap has been recognized in colormap design, where data-aware strategies are already commonly developed~\cite{Tominski08, Szafir18, Lu21, Reda18, Zeng22}. For example, Tominski~\etal~\cite{Tominski08} proposed histogram-based colormap enhancement for 2D scalar fields, while Zeng~\etal~\cite{Zeng22} leveraged spatial variations to guide continuous colormap adjustment. However, these approaches are typically tailored to specific design goals and do not provide a general formulation for assessing colormaps with respect to the underlying data.

We propose to address this gap by incorporating data directly into colormap assessment. Building upon the framework of Bujack~\etal~\cite{Bujack18}, we model this process as a composite mapping from the data manifold to perceptual color space. Starting from the perceptual metric in color space, we propagate local data variations through the chain rule to induce a data-aware perceptual metric on the data manifold~\cref{subsec:dataAssess}. Aggregating this local metric over the data domain yields data-aware counterparts of the established measures in~\cite{Bujack18}, including discriminative power, uniformity, and smoothness. For linear data, these reduce to their data-independent counterparts; for non-uniform data, they capture the joint effect of the colormap and the underlying data. The formulation also yields additional measures such as perceptual anisotropy and degeneracy.

We evaluate the proposed formulation from multiple aspects. To examine whether the data-aware measures remain consistent with established colormap knowledge when data influence is minimal, we first compare rankings with the data-independent measures of Bujack~\etal~\cite{Bujack18} under a linear field (D2), and then analyze how rankings shift as data vary across diverse datasets. We further test consistency with prior perceptual study by reproducing quantity estimation and gradient comparison tasks from Reda~\etal~\cite{Reda18} and comparing the trends observed in our measures with their empirical findings. To demonstrate generality, we extend the formulation from 1D to 2D colormaps and show how it can be integrated into histogram-equalization-based colormap optimization. An interactive system accompanies our approach, allowing users to explore colormap performance under varying data and evaluation criteria.

Our contributions are as follows:
\begin{itemize}
    % \item We introduce a data-aware formulation for colormap assessment by introducing a perceptual metric on the underlying data manifold, through which established measures such as discriminative power, uniformity, and smoothness become functions of the data rather than intrinsic properties of the colormap. The metric also enables additional quality measures, including perceptual anisotropy and degeneracy, that characterize previously unexamined aspects of colormap behavior.

    \item We introduce data-aware colormap assessment with a perceptual metric evaluated on the underlying data manifold. Established colormap measures such as discriminative power, uniformity, and smoothness become functions of the data rather than of the colormap alone. To ensure that data structures are maximally perceivable across all data values, we introduce perceptually uniformized variants of popular colormaps that have exactly uniform perceptual speed under a perceptual colorspace metric.

    \item We validate the proposed formulation through comprehensive evaluations, demonstrating that it recovers classical data-independent measures as a special case under linear field (D2) and yields assessments that better align with perceptual judgments when data are non-uniform.
    
    \item We extend the formulation to 2D colormap assessment and design, and develop an interactive system for data-aware exploration.

\end{itemize}

All source code, tested colormaps and datasets are publicly available at \url{https://osf.io/grzve/overview?view_only=1332a391dbd64c579ee1ff0e8e0f2de9}, along with an interactive system accessible at \url{https://101.201.80.169}.
\section{Related Work}

\subsection{Colormap Design for Scalar Visualization}

% 最后一句调整语序 early work 之前 [check]
% Effective colormap design ensures that perceptual color differences correspond meaningfully to differences in underlying scalar values, enabling users to accurately perceive gradients, boundaries, and structures in the data~\cite{zhou2016survey, ware2021, Bujack18}. 
% Early work established foundational design principles grounded in human perception. Ware~\etal~\cite{ware2021} proposed that luminance should vary monotonically with scalar values to prevent spurious visual boundaries. Rogowitz and Treinish~\cite{Rogowitz98} further observed that the luminance channel is best suited for encoding low-frequency features such as global gradients and large-scale structures, while hue is more effective for representing high-frequency local details. These principles remain widely adopted in colormap design today.

% Building on these foundations, subsequent research developed perceptually uniform colormaps as improved alternatives to traditional rainbow colormaps, which are known to introduce visual artifacts and misinterpretations.

Effective colormap design ensures that perceptual color differences correspond meaningfully to differences in underlying scalar values, enabling users to accurately perceive gradients, boundaries, and structures in the data~\cite{zhou2016survey, ware2021, Bujack18}. 
Many principles that remain widely adopted in colormap design today originate from early work grounded in human perception. 
Ware~\etal~\cite{ware2021} proposed that luminance should vary monotonically with scalar values to prevent spurious visual boundaries. 
Rogowitz and Treinish~\cite{Rogowitz98} further observed that the luminance channel is best suited for encoding low-frequency features such as global gradients and large-scale structures, while hue is more effective for representing high-frequency local details.

Building on these foundations, subsequent research developed perceptually uniform colormaps as improved alternatives to traditional rainbow colormaps, which are known to introduce visual artifacts and misinterpretations.
Borland and Taylor~\cite{Borland07} conducted controlled experiments confirming that the non-monotonic luminance profile of rainbow colormaps leads to missed pathological features in medical imaging. Borkin~\etal~\cite{borkin2011} incorporated eye-tracking into diagnostic tasks and showed that perceptually uniform colormaps substantially reduce misjudgment rates in artery stenosis detection. These findings motivated the development of standardized colormaps now integrated into major scientific visualization tools~\cite{Ware19, wyszecki2000}. Notable examples include the Viridis family by Van der Walt~\etal~\cite{vanderwalt2015}, which has become the default in the Python scientific computing ecosystem, and Moreland's cool-warm diverging colormap~\cite{Moreland09}, widely adopted in tools such as ParaView~\cite{Ayachit15} and VTK~\cite{vtkBook} for its perceptual uniformity and cross-device consistency.
The success of these colormaps reflects a broader consensus that colormap quality can and should be grounded in perceptual properties, motivating the development of formal quality measures discussed in \cref{subsec:colorassessment}.

\subsection{Quantitative Measures for Colormap Assessment}
\label{subsec:colorassessment}
To systematically compare colormaps, researchers have proposed quality measures that characterize the perceptual properties of color mappings. These measures are commonly defined in perceptually motivated color spaces such as CIELAB and CIELUV~\cite{wyszecki2000, Ware19}, where geometric distances approximate perceived color differences. Widely used measures include Just Noticeable Difference (JND) thresholds based on human visual sensitivity~\cite{Ware19}, perceptual gamut coverage, and the local uniformity of color-difference gradients~\cite{Reda19}.

Several approaches have been proposed for assessing continuous colormaps within this geometric setting.
Bujack~\etal~\cite{Bujack18} formalized the mathematical definitions of key colormap properties, including discriminative power, uniformity, and ordering, based on perceptual speed and mutual distances along the colormap. Nardini~\etal~\cite{nardini2020a} developed a standardized testing environment that enables reproducible quantitative comparison of colormaps under these measures. Their framework evaluates colormaps in the context of benchmark data, whereas our work complements these approaches by introducing a formulation that incorporates data distributions into the evaluation process.

\subsection{Data-Aware Colormap Design}
Data characteristics play an important role in colormap design for a long time. A central idea is that the colormap should be adapted to the data so that perceptually important color differences are allocated to where the data most needs them.

Early work by Tominski~\etal~\cite{Tominski08} proposed task-driven colormap transformations, including histogram equalization and Box-Whisker plot adaptation, to better match colormaps to data characteristics. Thompson~\etal~\cite{Thompson13} extended this idea by matching the slope of colormaps to the underlying volumetric data, observing that without such adaptation, small features become imperceptible due to the simultaneous contrast effect. Schulze-Wollgast~\etal~\cite{schulzewollgast2005} proposed an adaptive method that allocates a larger portion of the color gamut to data intervals with higher statistical significance based on metadata such as mean and median values. Mittelst\"adt~\etal~\cite{mittelstadt2014contrast, mittelstadt2015personalized} developed contrast compensation methods that adjust color encodings based on data and user-specific perception models to preserve discriminability in densely populated data regions. 
More recently, Zeng~\etal~\cite{Zeng22} formulated colormap adjustment as an optimization problem driven by spatial variations in the data field, and introduced a local variation dissimilarity measure to quantitatively evaluate whether the adjusted colormap better reveals spatial structures. This measure represents an early attempt to assess colormap effectiveness in a data-dependent manner, though it was designed specifically for their optimization pipeline rather than as a general-purpose quality measure.

These approaches collectively demonstrate that data meaningfully affect colormap effectiveness and that adapting colormap design to the data can improve perceptual outcomes. However, they focus on colormap generation or optimization, and the data characteristics that guide the design process are not incorporated into the colormap quality measures themselves. Our work addresses this gap by defining data-aware counterparts of established quality measures, enabling quantitative colormap assessment under specific data.
\section{Methodology}
In this section, we first briefly review the theoretical colormap evaluation framework proposed by Bujack~\etal~\cite{Bujack18}, covering both its fundamental measures and the quality measures defined therein. We then present our data-aware reformulation of continuous colormap, along with the redefinition of corresponding quality measures. 

\subsection{Background: Theoretical Colormap Assessment}
\label{subsec:Background}
The basic idea of Bujack~\etal's~\cite{Bujack18} formulation is to avoid the influence of data, tasks, and colorspace. Accordingly, they treat a continuous colormap as a curve in a 3D perceptual colorspace and leverage the perceptual distance between colors as a unified measure to mathematically formalize the most common design rules~\cite{Wainer80, Borland07, Silva07, Silva11, Bujack18, Bernard15}.

\subsubsection{Fundamental Colormap Measures}
\label{subsubsec:colormeasure}
In Bujack~\etal's~\cite{Bujack18} framework, a 1D continuous colormap is represented as a parametric curve $\gamma$ in a colorspace:
\begin{equation}
        \gamma:[0,1]\to C,
    \label{eq:colorcurve}
\end{equation}
where $C$ is the perceptual colorspace such as CIELAB~\cite{wyszecki2000}. 
Based on this definition, the authors define a \textbf{speed} measure, which depicts the rate of change between two colors along the colormap trajectory. 
Given a perceptual color distance metric $\Delta E:C\times C\to\mathbb R$, the speed $V$ between two colors at specific parametric positions $t_k$ and $t_i$ on the trajectory can be defined as:
\begin{equation}
    V(t_i,t_k)=\frac{\Delta E(t_i,t_k)}{t_k-t_i},
\end{equation}
where $\Delta E(t_i,t_j)$ is the shorthand of $\Delta E(\gamma(t_i),\gamma(t_j))$. This formulation mainly describes the relationship between any two colors in a colormap, known as the \textbf{global} measure. We can derive a \textbf{local} measure by simply considering neighboring pairs of colors by $v(t_i)= V(t_i,t_{i-1})$.
% check bujack v(t_i)
In this study, we mainly focus on the \textbf{local} measure, as perceptual sensitivity is inherently differential--observers respond to incremental changes between neighboring colors rather than absolute differences~\cite{Healey12, Reda18, ware2021}. 

\subsubsection{Colormap Quality Measures}
Within the geometric colormap quality measures, Bujack~\etal~\cite{Bujack18} have formulated measures for the most common colormap design rules, including discriminative power, uniformity, and smoothness. 

\vspace{.5em}
\noindent{\textbf{Local Discriminative Power}}. Discriminative power describes how well different colors can be distinguished. In Bujack~\etal~\cite{Bujack18}'s work, it is related to the average local speed along the colormap trajectory:
\begin{equation}
        D:=\frac{\sum_{i=1}^n v(t_i)}{n-1},
        \label{eq:gbudp}
\end{equation}
where $n$ indicates the number of color samples in a continuous colormap, therefore the average is taken over the $n-1$ intervals between adjacent. Larger values of $D$ indicate higher discriminative power.

\vspace{.5em}
\noindent{\textbf{Local Uniformity}}. Uniformity refers to whether equal steps in the parameter domain correspond to equal perceptual differences. It is characterized by the consistency of local speeds along the trajectory. This can be measured by the variance of local speed:
\begin{equation}
        U:=\mathrm{Var}\left[v(t_i)\right],
        \label{eq:gbuuniform}
\end{equation}
where lower values of $U$ indicate more uniform perceptual change.

\vspace{.5em}
\noindent{\textbf{Smoothness}}. Smoothness describes whether perceptual changes along the colormap occur without abrupt transitions. In Bujack~\etal's framework~\cite{Bujack18}, smoothness is treated as a design guideline rather than a quantitative measure, as it involves multiple factors including variations in local perceptual speed, geometric changes in the colormap trajectory, and differences in visual importance. It is therefore characterized qualitatively through properties such as continuity, low curvature, and the absence of perceptual boundaries. A rigorous formulation would require notions such as curvature, which depend on an inner product and are not well-defined under general perceptual metrics. Bujack~\etal~\cite{Bujack18} noted this difficulty and left the formulation of smoothness under non-Euclidean metrics as future work.

\subsubsection{Discussion}
In summary, the framework of Bujack~\etal~\cite{Bujack18} provides a principled formulation for colormap assessment by representing a colormap as a continuous curve in perceptual colorspace and evaluating it through perceived distances within a unified and data-independent formulation.
By relying solely on perceptual distances, the framework remains independent of both the choice of colorspace and the underlying distance metric.
A key strength of this formulation is its deliberate independence from any specific dataset or task, with all measures defined purely in terms of the trajectory $\gamma$ in perceptual space, effectively corresponding to the assumption of a linear varying scalar field. 

At the same time, this abstraction also highlights a limitation in applicability: \emph{in real-world visualization, perceptual variation arises not only from the colormap itself but also from the underlying data}~\cite{Moreland09, Bujack18, Bujack26}.
While the data-independent formulation is essential for generality, it does not explicitly account for how colormaps interact with data characteristics--a topic extensively studied in existing colormap design research~\cite{Reda18, Reda20}.
Building on the geometric insight introduced by Bujack~\etal~\cite{Bujack18}, we extend this perspective by viewing visualization as a mapping from the data manifold to perceptual colorspace, in which perceptual variation is jointly determined by both the colormap and the underlying data (\cref{subsec:dataAssess}).
% + Bujack 2026 ref
% composition mapping
% distribution 问题

\subsection{Data-Aware Colormap Assessment}
\label{subsec:dataAssess}

% metric 在第一次出现的地方提一下
% 缺了一层
% spatial domain 看起来很怪
In this section, we develop a data-aware formulation building on the foundational definitions of Bujack~\etal~\cite{Bujack18} (\cref{subsec:Background}). The key idea is to analyze the visualization formed by color-encoded data through the colormapping process, rather than the colormap alone. To this end, we first formulate the colormapping process as a geometric composition of the data field, the colormap, and the perceptual metric (\cref{subsubsec:geometric}). We then introduce the data-aware perceptual metric on the data domain based on the chain-rule decomposition of this composition (\cref{subsubsec:tensor}). Building on this metric, we derive fundamental data-aware colormap measures in \cref{subsubsec:fundamental}.

\subsubsection{Geometric Colormapping Formulation}
\label{subsubsec:geometric}
% We formulate the colormapping process as a composition of the data field, the colormap, and the perceptual metric, so that local perceptual effects of the colormapping process can be analyzed.

\vspace{0.5em}
\noindent\textbf{Data Manifold}. 
Let $M$ denote the spatial domain on which the scalar field $f$ is defined, modeled as a two-dimensional smooth manifold. We refer to $M$ as the \emph{data manifold}. The distribution of $f$'s values across $M$ determines which regions of the colormap are applied in a given visualization, connecting the data discussed earlier to the geometric setting introduced here. Formally, $f$ is a smooth map:
\begin{equation}
    f: M \to W,
\end{equation}
where $W \subset \mathbb{R}^k$ is the value space. When $k=1$, $f$ is a scalar field, while when $k>1$, it is a bivariate or more general multi-attribute field.

\vspace{0.5em}
\noindent\textbf{Geometric Colormap Representation}. 
Furthermore, we extend the colormap definition in \cref{eq:colorcurve} to a dimension-independent form:
\begin{equation}
        \gamma: W \to C.
\end{equation}
To quantify perceptual differences in color space, we follow the assumption that color space $C$ is locally Riemannian~\cite{Pant10, Molchanov25}, equipped with a metric $h$ that encodes how color differences are perceived in different directions:
\begin{equation}
        h = \sum_{\mu}\sum_{\nu} \underbrace{h_{\mu\nu}(c)}_{H\text{'s elem}}\, \mathrm{d}c^\mu \otimes \mathrm{d}c^\nu,
\end{equation}
where $H$ is the matrix representation of $h$. The value of matrix element $h_{\mu\nu}(c)$ is determined by the choice of $\Delta E$. For example, $\Delta E_{76}$ makes $h$ a Euclidean metric, so that $H$ can be treated as a constant matrix, while $\Delta E_{2000}$ makes $h$ a non-Euclidean metric, so that $H$ is given by a space-varying local quadratic form approximation.

\vspace{0.5em}
\noindent\textbf{Color Mapping Definition}.
The color-encoded scalar field visualization is defined as the composite mapping:
\begin{equation}
        \Phi := \gamma \circ f,\quad \Phi : M \to C .
        \label{eq:vismap}
\end{equation}
which maps points in $M$ to colors in perceptual color space and serves as the basis for constructing the data-aware perceptual metric.

\begin{figure}[t!]
    \centering
    \includegraphics[width=.9\linewidth]{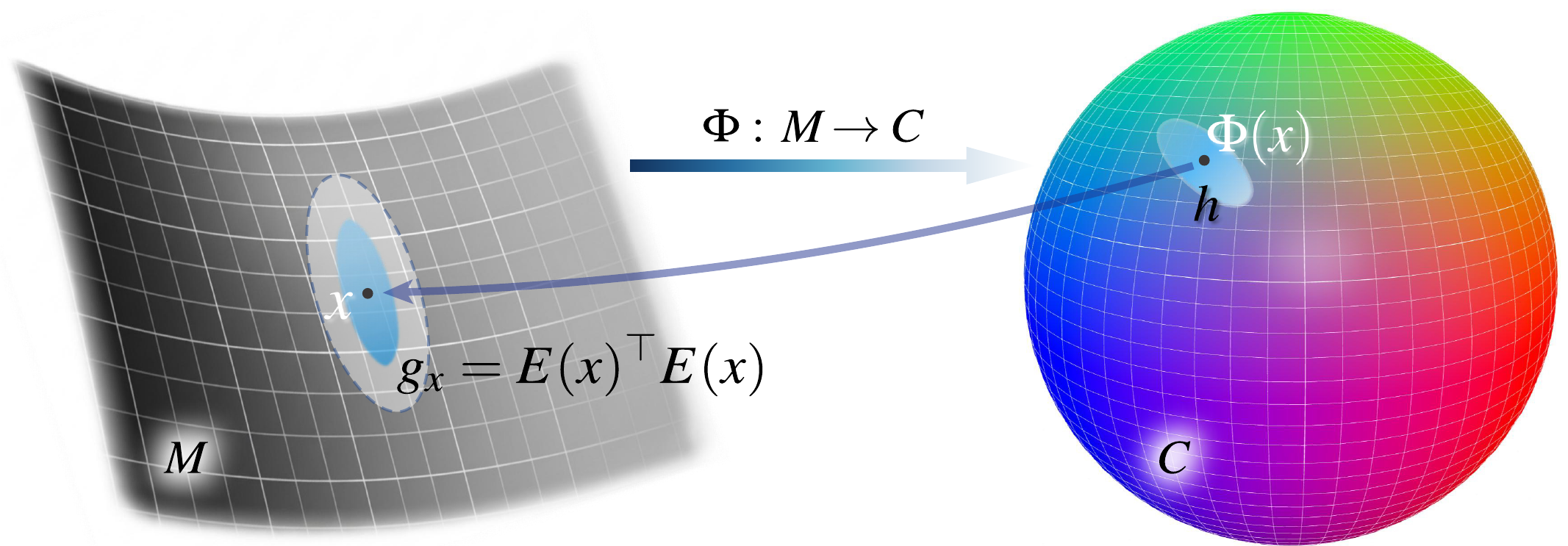}
    % \vspace{-0.5em}  
    \caption{\textbf{Data-aware perceptual metric.} The visualization map $\Phi:M\to C$ sends each data-domain point $x$ to a color $\Phi(x)$ in perceptual color space, where perceptual color differences are measured by the metric $h$. Through the chain-rule decomposition~\cref{eq:chainrule}, the local perceptual derivatives $E(x)$ of $\Phi$ pulls $h$ back to a metric $g_x = E(x)^\top E(x)$ on the data domain, capturing how local data variations around $x$ translate into perceived color differences.}
    \label{fig:pullback}
    \vspace{-0.5em}
\end{figure}

\subsubsection{Data-Aware Perceptual Metric}
\label{subsubsec:tensor}
Following the perspective introduced in~\cref{sec:Introduction}, a colormap should be assessed not in isolation, but by how it transforms variations in a specific data manifold into perceptual color differences. We therefore analyze how infinitesimal changes on the data manifold propagate through the composed visualization map $\Phi$. 

For $x \in M$ and a tangent vector $u \in T_xM$, the chain rule decomposes the color-space variation as $\mathrm{d}\Phi_x(u)=\mathrm{d}\gamma_{f(x)}(\mathrm{d}f_x(u))$. In local coordinates:
\begin{equation}
    J_\Phi(x)
    =
    J_\gamma(f(x))\, J_f(x),
\end{equation}
where $J_f$ and $J_\gamma$ denote the Jacobians of $f$ and $\gamma$, respectively.

To convert this color-space variation into perceptual variation, we apply the Riemannian metric $h$ on $C$. Let $H(c)$ be the matrix representation of $h$ at color $c$, with Cholesky factorization $ H(c)=L(c)^\top L(c)$. Under this factorization, the perceptual effect of a spatial tangent vector $u$ reduces to an ordinary Euclidean norm: $\Vert L(\Phi(x))J_\Phi(x)u\Vert$. We define the local perceptual derivative of the visualization map as:
\begin{equation}
E(x)
:=
L(\Phi(x))J_\Phi(x)
=
\underbrace{L(\gamma(f(x)))}_{\text{perception}}
\underbrace{J_\gamma(f(x))}_{\text{colormap}}
\underbrace{J_f(x)}_{\text{data}}.
\label{eq:chainrule}
\end{equation}
The data-aware perceptual metric at $x$ is then:
\begin{equation}
    g_x=E(x)^\top E(x).
\end{equation}
$g_x$ is a symmetric positive semidefinite 2-tensor that encodes, at every point and in every direction on the data manifold, how strongly a local data variation translates into a perceived color difference. All subsequent measures \cref{subsubsec:Data-Aware Colormap Quality Measures} are derived from $g_x$. Note that \cref{eq:chainrule} is closely related to classical multi-channel image gradient representations~\cite{Zenzo86}; the key difference is that we apply the differential structure to the composed visualization map for perceptual colormap assessment.

\begin{figure}[t!]
    \centering
    \includegraphics[width=1.0\linewidth]{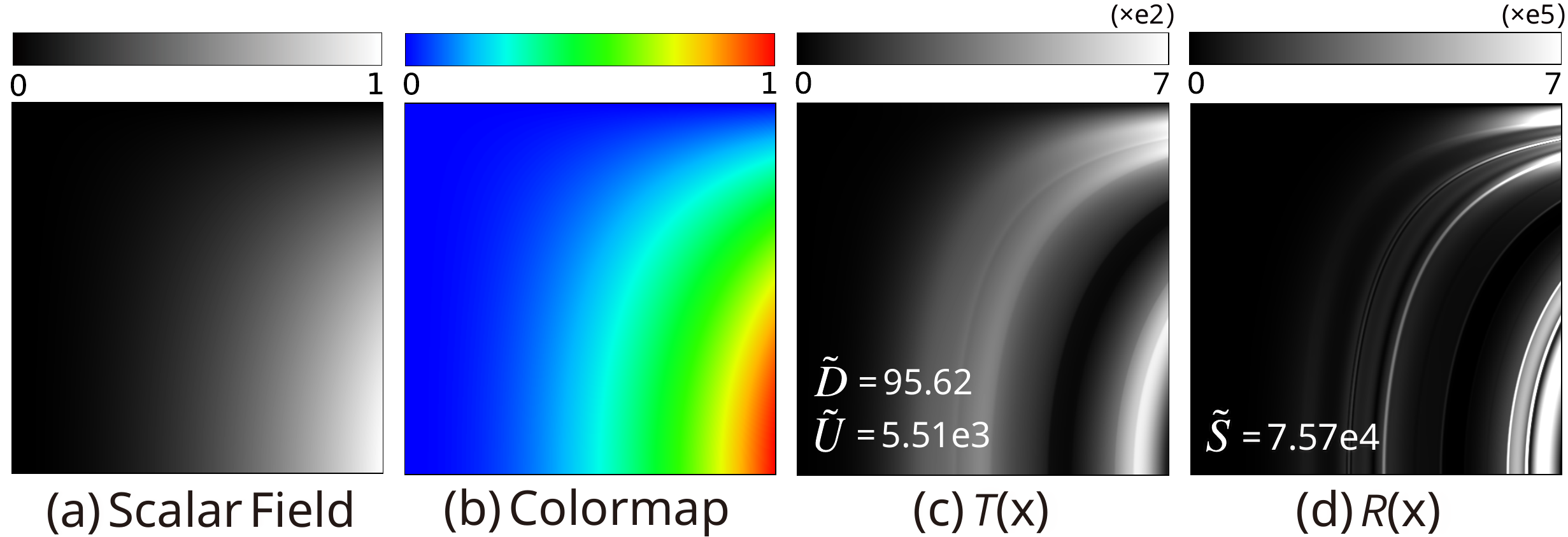}
    \caption{\textbf{An illustration of $T(x)$ and $R(x)$}. (a) A synthetic \textit{gradient variation field} dataset D3, followed by the visualization effects on: (b) The dataset encoded with \textit{rainbow}. (c) $T(x)$, in which the white indicates values with higher local perceptual response. (d) $R(x)$, in which the white indicates values with intense perceptual change.}
    \label{fig:illustration}
\end{figure}

\subsubsection{Fundamental Data-Aware Colormap Measures}
\label{subsubsec:fundamental}

In analogy to the speed $v(t)$ introduced in \cref{subsec:Background}, which captures
perceptual change along a 1D colormap trajectory, we use $g_x$ to define
a scalar summary of local perceptual change~\cite{Carmo92, Lee18}. The trace of $g_x$ sums the perceptual changes along each coordinate direction, and equals the squared Frobenius norm of $E(x)$:
\begin{equation}
    T(x) :=\sqrt{\mathrm{tr}(g_x)}=\Vert E(x) \Vert_F,
    \label{eq:tr}
\end{equation}
As shown in~\cref{fig:illustration}c, a larger $T(x)$ indicates stronger perceived color changes around $x$ across all spatial directions. For a scalar field $f:M\to \mathbb{R}$ with a 1D colormap $\gamma(t)$, we have $J_f(x)=\nabla f(x)^\top$ and $J_\gamma(f(x))=\gamma'(f(x))$. Thus, \cref{eq:chainrule} reduces to a rank-one form, and taking its Frobenius norm gives:
\begin{equation}
    T(x)
    =
    \sqrt{\tilde g(f(x))}
    \,
    \Vert \nabla f(x) \Vert .
    \label{eq:tr1d}
\end{equation}
where $\tilde g(t)=\gamma'(t)^\top H(\gamma(t))\gamma'(t)$ is the squared local perceptual speed of the colormap. The perceptual response at each point is thus jointly determined by the colormap's local perceptual speed and the data gradient. In the simplest case of a Euclidean metric such as $\Delta E_{76}$ and a field $f$ mapping to $[0,1]$ with a constant-magnitude gradient, $T(x)$ becomes $|\gamma'(f(x))|$, which is proportional to the classical speed $v(f(x))$ of Bujack~\etal~\cite{Bujack18}.
% where $\tilde g(t)=\gamma'(t)^\top H(\gamma(t))\gamma'(t)$ is the squared local perceptual speed of the colormap. The perceptual response at each point is thus jointly determined by the colormap's local perceptual speed and the data gradient. In the simplest case of a Euclidean metric such as $\Delta E_{76}$ and a unit-gradient linear field $f$, $T(x)$ becomes $|\gamma'(f(x))| = v(f(x))$, where $f(x)$ is normalized to $[0,1$], recovering the classical speed of Bujack~\etal~\cite{Bujack18}.

\subsection{Data-Aware Colormap Quality Measures}
\label{subsec:measure}
Building on the data-aware perceptual metric $g_x$, we reformulate the quality measures of Bujack~\etal~\cite{Bujack18}--covering discriminative power, uniformity, and smoothness--for data-aware colormap assessment (\cref{subsubsec:Data-Aware Colormap Quality Measures}). 
The original formulation evaluates these measures along the colormap alone; ours evaluates them over the data manifold through the spatially varying metric $g_x$.
We also extend the formulation to anisotropy and degeneracy (\cref{subsubsec:Extensions to Other Colormap Quality Measures}).

\subsubsection{Data-Aware Colormap Quality Measures}
\label{subsubsec:Data-Aware Colormap Quality Measures}
\noindent\textbf{Data-Aware Discriminative Power}. 
Data-aware local discriminative power quantifies how strongly small variations in the data field affect perception across a region. In analogy to \cref{eq:gbudp}, it is computed by averaging $T(x)$ over $\Sigma \subset M$:
\begin{equation}
    \tilde D := \frac{1}{|\Sigma|} \int_{\Sigma} T(x)\, \mathrm{d}\sigma,
    \label{eq:dp}
\end{equation}
where $\mathrm{d}\sigma$ denotes a small area element on $\Sigma$. The larger values of $\tilde D$ indicate stronger perceptual responses.

\vspace{0.5em}
\noindent\textbf{Data-Aware Local Uniformity}. 
Data-aware local uniformity measures how evenly the visualization distributes perceptual differences across a region. Analogy to \cref{eq:gbuuniform}, it is defined as the variance of $T(x)$ over $\Sigma$:
\begin{equation}
    \tilde U := \mathrm{Var}_{x\in\Sigma}\left(T(x)\right),
    \label{eq:uniform}
\end{equation}
where lower values of $\tilde U$ indicate a more uniform perceptual response.

\vspace{0.5em}
\noindent\textbf{Data-Aware Local Smoothness}. 
Smoothness describes how gradually perceptual response varies across the data field. As discussed in \cref{subsec:Background}, smoothness was identified as a design guideline but not given a formal measure in prior work. The data-aware metric $g_x$ enables such a definition: we quantify local roughness as the spatial variation of $g_x$:
\begin{equation}
    R(x) := \Vert \nabla g_x \Vert_F.
\end{equation}
Intuitively, $R(x)$ captures how rapidly the perceptual response changes between neighboring locations: high values indicate abrupt perceptual transitions, such as yellow-to-green and green-to-cyan transitions of \textit{rainbow}, while low values indicate smooth ones, see \cref{fig:illustration}d. The data-aware smoothness over a region $\Sigma$ is then defined as the average roughness:
\begin{equation}
    \tilde S := \frac{1}{|\Sigma|} \int_\Sigma R(x)\, \mathrm{d}\sigma.
\end{equation}

\subsubsection{Extensions to Other Colormap Quality Measures}
\label{subsubsec:Extensions to Other Colormap Quality Measures}
\vspace{0.5em}
\noindent\textbf{Data-Aware Perceptual Anisotropy}.
Anisotropy~\cite{Molchanov25} measures how perceptual sensitivity varies across directions at a given point in the data manifold. A natural scalar summary of directional imbalance is:
\begin{equation}
        % \Lambda(x) = \frac{\lambda_1(x) - \lambda_2(x)}{\lambda_1(x) + \lambda_2(x)},
\Lambda(x)
=
\frac{\lambda_1(x)-\lambda_2(x)}
{\lambda_1(x)+\lambda_2(x)},
\end{equation}
where $\lambda_1(x)\ge \lambda_2(x)$ are the eigenvalues of $g_x$.
Larger values of $\Lambda(x)$ indicate a stronger directional bias in the perceptual response. The data-aware anisotropy over a region $\Sigma \subset M$ is defined by averaging:
\begin{equation}
    \begin{split}
        \tilde A:=\frac{1}{|\Sigma|} \int_\Sigma \Lambda(x) \, \mathrm{d}\sigma.
    \end{split}
\end{equation}
For scalar fields with 1D colormaps, $E(x)$ is rank-deficient and all perceptual variation aligns with the data gradient, so anisotropy carries no additional information beyond $T(x)$. For multi-dimensional colormaps, however, the eigenstructure of $g_x$ reveals directional imbalances in perceptual sensitivity that are inaccessible to scalar measures alone. Note that this data-aware anisotropy differs from the intrinsic anisotropy studied by Molchanov~\etal~\cite{Molchanov25}, which is a property of the perceptual color space itself. Here, $\Lambda(x)$ is derived from $g_x$ and therefore reflects the joint effect of the colormap and the local data field on directional perceptual sensitivity.

\begin{figure}[t!]
    \centering
    \includegraphics[width=1.0\linewidth]{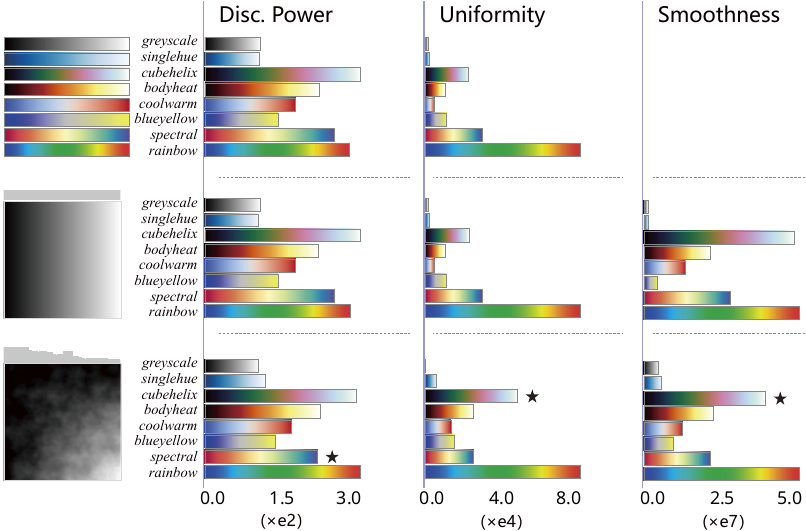}
    \caption{\textbf{Comparison between data-independent and data-aware measures}. For each colormap, the bar plots show discriminative power, uniformity, and smoothness (data-aware only) for a \textit{linear field} (D2) and a \textit{Perlin noise field} (D1). For the linear field, the data-aware results closely follow the data-independent baselines. $\bigstar$ is used to highlight the values that changed the most for D1, compared to the baseline.}
    \label{fig:exp-comparegbu}
\end{figure}
\vspace{0.5em}
\noindent\textbf{Data-Aware Perceptual Degeneracy}. 
To detect whether perceptual variation is preserved across all directions, we consider the determinant of $g_x$~\cite{Harris88, Weickert91}:
\begin{equation}
    G(x):=\det(g_x)=\det\big(E(x)^\top E(x)\big).
\end{equation}
which serves as a local indicator of degeneracy. Small values of $G(x)$indicate that the perceptual response is nearly degenerate, meaning that perceptual variation is lost in at least one direction, while larger values indicate that variation is maintained across directions. In contrast to $T(x)$, which sums perceptual responses across directions, $G(x)$ captures whether responses exist jointly in all directions. The data-aware degeneracy measure over a region $\Sigma \subset M$ is:
\begin{equation}
    \tilde G := \frac{1}{|\Sigma|} \int_{\Sigma} G(x)\, \mathrm{d}\sigma.
\end{equation}
We illustrate the role of these two measures through two 2D colormaps in~\cref{subsec:generalization}, where the full eigenstructure of $g_x$ becomes meaningful. 

\begin{figure*}[t!]
    \centering
    \includegraphics[width=1.0\linewidth]{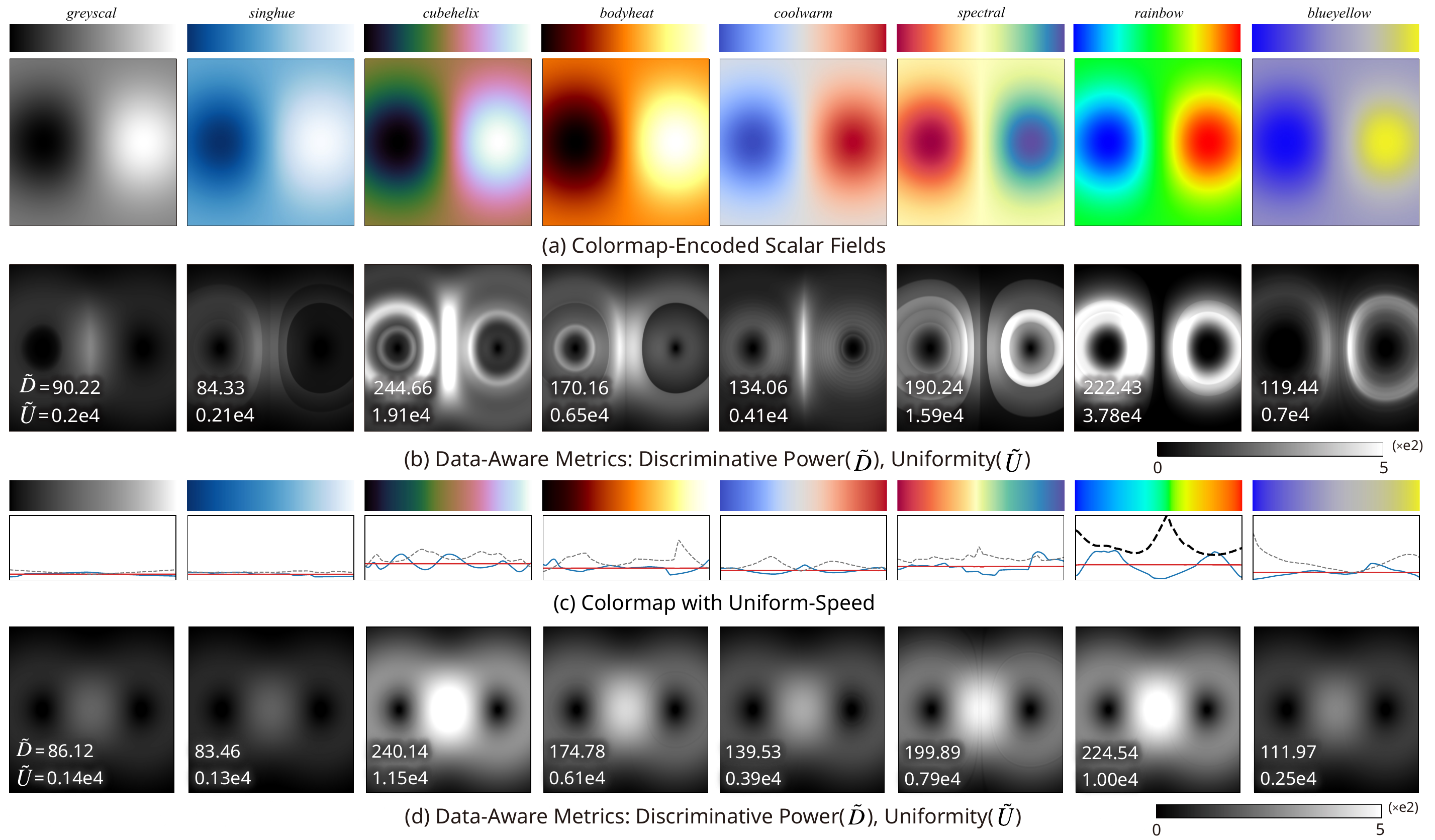}
    \caption{\textbf{Data-aware perceptual analysis on a symmetric two-Gaussian field (D4)}. (a) Color-encoded scalar fields under different colormaps. (b) Corresponding local fields $T(x)$ with data-aware discriminative power $\tilde D$, uniformity $\tilde U$, and smoothness $\tilde S$, showing how perceptual emphasis is distributed across the data manifold. (c) Colormap speed: the original (blue \legendline{cyan}), uniformized under $\Delta E_{2000}$ (red \legendline{red}), and re-evaluated under $\Delta E_{76}$ (black dashed \legendline{black, dashed}, least uniform case \textbf{highlighted}).
    (d) $T(x)$ and data-aware measures ($\tilde D$, $\tilde U$) under the uniformized colormap.}
    \label{fig:exp-spatialanalysis}
\end{figure*}

\section{Results}
\subsection{Experimental Setup}
\vspace{.5em}
\noindent\textbf{Data}.
We selected nine datasets for evaluating the proposed data-aware colormap assessment, comprising five synthetic datasets and four real-world scientific datasets. The synthetic datasets include \emph{Perlin noise field}~\cite{Perlin85} (D1), \emph{linear field} along $x$-axis or $y$-axis (D2), \emph{gradient variation field}~\cite{nardini2020a} (D3), \emph{gaussian field} $N(x\mid\mu,\sigma^2)$ (D4), and \emph{ripples} from Langermann function~\cite{nardini2020a} (D5). The real-world scientific datasets include \emph{ocean salinity} simulated with the high-resolution
global ocean circulation model MPIOM~\cite{Jungclaus13} (D6), \emph{temperature} (D7), \emph{sea surface temperature} (D8), \emph{ocean salinity} (D9). 
These datasets cover a broad diversity of distributions, reducing potential biases introduced by any single data characteristic. 

\vspace{0.5em}
\noindent\textbf{Colormap}. 
We selected eight continuous colormaps spanning a range of luminance and chroma properties~\cite{Reda18, Zeng22}: one baseline (\textit{gray}), two sequential colormaps with monotonically increasing luminance over a limited number of hues (\textit{singlehue}, \textit{bodyheat}), one spiral colormap with monotonically increasing luminance and multiple hues (\textit{cubehelix}), two diverging colormaps with uniformly-stepped luminance (\textit{coolwarm}, \textit{spectral}), one diverging colormap with uniformly-stepped saturation (\textit{blueyellow}), and one fully saturated multi-hue colormap (\textit{rainbow}). This selection covers the major colormap categories and includes both widely recommended and commonly criticized designs, allowing the data-aware measures to be tested across diverse perceptual profiles.

\vspace{0.5em}
\noindent\textbf{Evaluation Setup.}
All scalar fields are normalized to $[0,1]$ and mapped to colors via continuous colormaps in CIELAB color space. Color differences are computed using $\Delta E_{76}$, $\Delta E_{94}$, and $\Delta E_{2000}$; we use $\Delta E_{2000}$ to illustrate results throughout the paper, with results for the other two formulas provided in the supplementary material. All data-aware measures are computed from the $g_x$ as described in \cref{subsec:dataAssess}, and can be evaluated globally or within user-defined regions of interest. Data-independent baselines are obtained from the same colormaps using the method of Bujack~\etal~\cite{Bujack18}. The evaluation pipeline is implemented in \texttt{Python} using \texttt{NumPy} and \texttt{Matplotlib}. Implementation details and complexity analysis are provided in the supplementary material.

\subsection{Quantitative Comparison}
\label{subsec:quantitative-comparison}
A key property of our data-aware formulation is that it generalizes the data-independent framework of Bujack~\etal~\cite{Bujack18}: under uniform data, the data-aware measures should recover the classical definitions. We first verify this by computing both the data-independent measures ($D$, $U$) and the data-aware measures ($\tilde D$, $\tilde U$, $\tilde S$) for all eight colormaps on a linear scalar field $f(x,y)=x$ (D2), which produces a uniform value distribution. As shown in the middle row of \cref{fig:exp-comparegbu}, the data-aware rankings closely match the data-independent baselines across all colormaps, confirming that the formulation reduces to the classical definitions when the data exerts no distributional bias.

\noindent We then apply all colormaps to datasets with non-uniform distributions. \cref{fig:exp-comparegbu} shows the results on a Perlin noise dataset with a skewed distribution (histogram shown at top). Compared to the uniform case, the data-aware rankings shift notably. For example, \textit{spectral}, which has the second highest discriminative power under the data-independent assessment, drops considerably when evaluated on this skewed distribution. Similarly, \textit{cubehelix} shows decreased uniformity and smoothness compared to its data-independent scores, indicating that its perceptual response becomes less even when the data concentrates in a narrow value range. These shifts illustrate that a colormap's effective perceptual quality on a given dataset may differ from its intrinsic quality as measured by data-independent metrics. The color-encoded visualizations in \cref{fig:teaser} provide a visual confirmation of these numerical differences. Similar patterns hold for other datasets and measures (see supplementary material for complete results).

\subsection{Spatial Analysis of the Perceptual Field}
\label{subsec:spatial-analysis}
Beyond global rankings, the $T(x)$ field provides spatially resolved information about how colormaps allocate perceptual emphasis across the data manifold. To illustrate this, we analyze the spatial structure of $T(x)$ on a synthetic test field (D4) constructed as the difference of two Gaussian functions, where $\mu=0.55$, $\sigma_1=\sigma_2=0.65$. This field contains smooth regions, localized high-gradient bands, and a symmetric zero-crossing boundary, making it a suitable probe for examining how different colormaps distribute perceptual resources in space.
We apply all test colormaps to this field and present the color-encoded visualizations alongside their $T(x)$ fields in \cref{fig:exp-spatialanalysis}. We also compute a speed-uniformized reparameterization for each colormap and apply the results to the same field, enabling direct comparison between original and uniform-speed settings. Differences between colormaps that are not readily apparent in the color-encoded visualizations become clearly visible in the $T(x)$ fields. We identify several characteristic patterns in how colormaps distribute perceptual response over the data manifold:
\begin{itemize}
    \item \textbf{Focused response at the structural boundary.} \textit{Coolwarm} shows high perceptual response around the zero-crossing, forming a narrow band aligned with the main structural boundary. Its perceptual colormap speed $\sqrt{\tilde g(t)}$ peaks near the midpoint of the value range, producing a clean central ridge in $T(x)$.
    \item \textbf{Distributed response across multiple structures.} \textit{Spectral}, \textit{rainbow}, and \textit{blueyellow} produce multiple high-response bands in $T(x)$, because their $\sqrt{\tilde{g}(t)}$ is elevated over several value intervals rather than concentrated at one. Perceptual emphasis is spread across multiple structures instead of focusing on the central transition. For instance, although \textit{spectral} is a diverging colormap like \textit{coolwarm}, its $T(x)$ field shows several uniform bands that dilute the emphasis on the zero-crossing.
    \item \textbf{Diverging-like behavior from sequential colormaps.} \textit{Cubehelix} and \textit{bodyheat} are sequential colormaps that nonetheless produce strong responses in multiple regions due to their non-uniform perceptual speed. \textit{Cubehelix} is a typical example: its helical trajectory in color space creates elevated perceptual sensitivity at several value intervals, producing both central and peripheral responses that resemble a diverging pattern in $T(x)$.   
\end{itemize}

\begin{figure}[t!]
    \centering
    \includegraphics[width=1.0\linewidth]{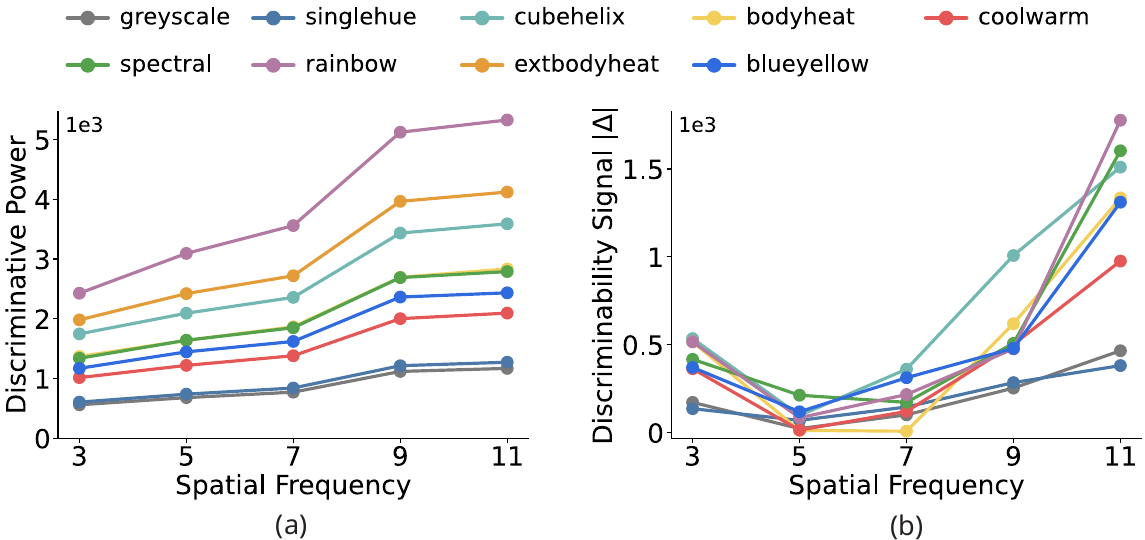}
     \caption{\textbf{Comparison with perceptual studies}. Values under different spatial frequencies: (a) Quantity estimation task, where a higher value indicates higher discriminative power. (b) Gradient comparison task, where a higher value indicates a higher discrimination signal.}
    \label{fig:exp-perception}
\end{figure}

These patterns illustrate that colormap category alone—sequential, diverging, or spiral—does not determine how perceptual emphasis is distributed on a given dataset. Even among diverging colormaps, the spatial allocation of perceptual response can vary substantially; conversely, sequential colormaps may exhibit diverging-like behavior when their perceptual speed is non-uniform.

The chain-rule factorization $T(x)=\sqrt{\tilde g(f(x))}\|\nabla f(x)\|$ explains why. As shown in~\cref{fig:exp-spatialanalysis}d, even under uniform-speed colormaps, $T(x)$ still exhibits spatial structure: the data gradient $\|\nabla f(x)\|$ and the perceptual colormap speed $\sqrt{\tilde g(t)}$ jointly shape the responses. Asymmetry arises when $\sqrt{\tilde g(t)}$ is not symmetric over the value range, and data gradient $\|\nabla f(x)\|$ further modulates where this imbalance becomes visually salient.
The choice of perceptual color metric also matters. \cref{fig:exp-spatialanalysis}c compares colormap speed under $\Delta E_{2000}$ and Lab-Euclidean $\Delta E_{76}$, showing that the same colormap can exhibit markedly different speed patterns depending on the metric.

The data-aware $T(x)$ makes these effects directly visible, revealing where perceptual resources are allocated and whether this allocation aligns with the data structure. In \cref{subsubsec:case2}, we further demonstrate that this spatial information can guide colormap adjustment to redistribute perceptual emphasis.

\subsection{Comparison with Perceptual Studies}
To validate the proposed data-aware measures, we replicate two perceptual tasks from Reda~\etal~\cite{Reda18}, which examine how colormap design and spatial frequency affect human judgment of scalar fields. We focus on their quantity estimation and gradient comparison experiments, as the two tasks most directly correspond to the discriminative power and gradient perception aspects of our formulation. Ideally, we expect our data-aware discriminative power $\tilde D$ to be negatively correlated with quantity estimation error, and the gradient discrimination signal $\Delta = \langle T\rangle_{\text{hi}} - \langle T\rangle_{\text{lo}}$ to be positively correlated with accuracy in gradient comparison. Following~\cite{Reda18}, we synthesize data using Perlin noise with five octaves and optimize the base frequency to achieve target median spatial frequencies. Since exact colormap definitions are not available for all colormaps used in Reda~\etal~\cite{Reda18}, we recover the \textit{blueyellow} colormap using the learning-based extraction method of Liu~\etal~\cite{Liu25}. More details can be found in the supplementary material.

\noindent{\textbf{Experiment 1: Quantity Estimation}}.
We simulate the quantity estimation task using $\tilde D$ (Eq.~\ref{eq:dp}) over procedurally generated terrain fields following \cite{Reda18}. As shown in~\cref{fig:exp-perception}a, the resulting ranking of colormaps is broadly consistent with the human error ranking reported by Reda~\etal~\cite{Reda18}--both increase with spatial frequency. Specifically, it is consistent with prior observations that higher spatial frequencies introduce larger local gradients, increasing both the perceptual differences captured by $\tilde D$ and the estimation errors observed in human studies. Despite this shared trend, the relative ordering across colormaps remains stable, reflecting intrinsic differences in perceptual sensitivity.

\noindent{\textbf{Experiment 2: Gradient Comparison}}.
We approximate the gradient comparison task by measuring $\Delta$ between high- and low-gradient regions across multiple terrain realizations. As shown in~\cref{fig:exp-perception}b, $|\Delta|$ provides a useful proxy for differentiating colormaps at high spatial frequencies, broadly consistent with trends reported in human performance where only certain colormaps remain discriminative under challenging conditions. At low spatial frequencies, all colormaps yield similar $|\Delta|$ values; at high frequencies, colormaps with rich hue variation or diverging designs produce larger $\Delta$ values, consistent with the empirical findings reported by Reda~\etal~\cite{Reda18}.

\noindent{\textbf{Observations}}. The observed consistency with human studies supports the hypothesis that the proposed measures capture aspects of perceptual behavior relevant to colormap assessment. From a geometric perspective, this consistency can be understood through the factorization $T= \sqrt{\tilde g(u)}\|\nabla u\|$, which separates colormap sensitivity from data variation. While not a substitute for direct perceptual validation, these results suggest that aggregating local $T(x)$ field quantities yields global assessments that align with empirical observations.

\begin{figure}[t!]
    \centering
    \vspace{-0.35cm}
    \includegraphics[width=1.0\linewidth]{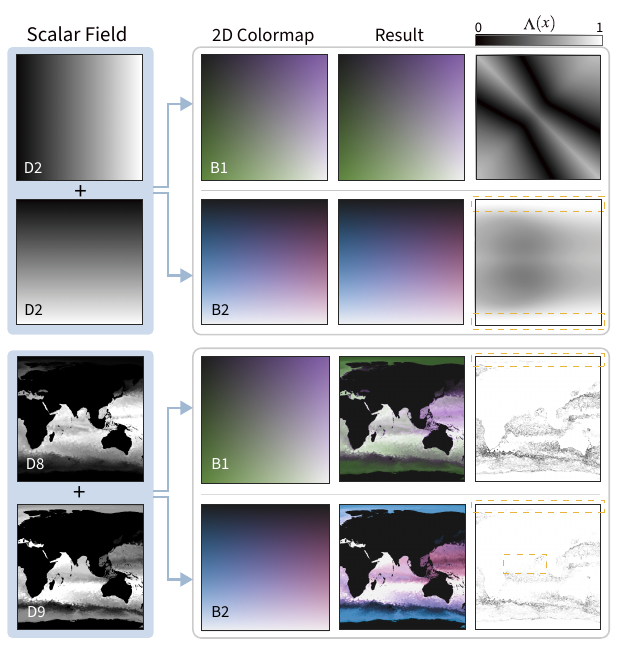}
    \caption{\textbf{2D colormap assessment.}
    Two pairs of scalar fields, including linear fields (D2) and real-world datasets (D8, D9), are mapped with two bivariate colormaps B1 and B2 from~\cite{Zhang06}. The $\Lambda(x)$ fields show that B1 yields more balanced directional sensitivity, while the cylindrical map B2 produces stronger directional bias near chroma-collapsed regions. Lost-direction areas are \textbf{highlighted} by yellow dashed boxes.}
    \label{fig:exp-bivariate}
\end{figure}

\subsection{Generalization}
\label{subsec:generalization}
\vspace{.5em}
\noindent\textbf{Generalization to 2D Colormap Assessment.}
The measures developed so far apply to scalar fields encoded with 1D colormaps, where $g_x$ is rank-deficient and $T(x)$-based quantities suffice to characterize local perceptual change. For bivariate fields encoded with 2D colormaps, $g_x$ can become full rank, and its eigenstructure reveals how perceptual sensitivity is distributed across directions. In this setting, the anisotropy $\Lambda(x)$ and the degeneracy indicator $G(x)$ become essential: $G(x)$ detects whether perceptual variation is preserved in both directions, while $\Lambda(x)$ indicates whether that variation is balanced or directionally biased.

\cref{fig:exp-bivariate} illustrates this extension of bivariate colormap on two synthetic linear fields (D2) and two real-world datasets (D8, D9). The observed patterns are consistent with the findings of Zhang and Healey~\cite{Zhang06}, who reported that constant-hue-plane or square-like colormap constructions generally produce more effective renderings than cylinder-based schemes, which are more prone to mutual interference between the two encoded variables. Our formulation provides a geometric explanation for these empirical observations: high $\Lambda(x)$ indicates that local sensitivity is dominated by one variable rather than shared between both. Due to space limitations, the corresponding $G(x)$ fields are provided in the supplementary material, showing that low $G(x)$ identifies regions where perceptual variation along one direction is effectively lost, corresponding to the interference reported in cylinder-based designs.

More broadly, the design of interpretable bivariate colormaps has long been guided by Trumbo's classical principles~\cite {Trumbo1981}--order, separation, rows and columns, and diagonal structure--which are rooted in the perceptual salience of luminance and opponent-color channels in color space. Consistent with these guidelines, prior studies~\cite {Zhang06, Kerren20} have further confirmed that effective 2D colormaps should balance contributions of both variables and avoid strong directional dominance. Our formulation provides a spatially resolved, quantitative means to diagnose how well these properties hold when a colormap is applied to a specific dataset. In particular, these design principles can be related to geometric constraints on the derived measures: maintaining $G(x) > 0$ over meaningful regions helps preserve that both variable directions remain perceptually encoded, aligning with the rows-and-columns principle; keeping $\Lambda(x)$ moderate prevents one variable from overwhelming the other; and the magnitude of local perceptual response further reflects the degree of separation between adjacent data values. Notably, the order principle relies on signed monotonic information along perceptual channels, which is not fully captured by the quadratic metric tensor, as it encodes only the magnitude of perceptual change rather than its directional sign.

\vspace{.5em}
\noindent\textbf{Generalization to Colormap Adjustment.}
\label{subsubsec:case2}
Beyond assessment, the proposed formulation can also inform colormap design. We demonstrate this through a histogram-equalization-based continuous colormap adjustment method (\cref{alg:hist}) that operates directly on the local perceptual field $T(x)$. 

The intended goal is to support comparison of spatial variation patterns across value ranges, including boundary salience, weak gradients, and subtle oscillatory structures~\cite{Reda18}. Standard value-histogram equalization tends to allocate more colors to high-density ranges, which may overemphasize noise or reduce the visibility of weak structures elsewhere~\cite{Zeng22}. In contrast, our method uses the perception histogram as a response profile and applies a controlled inverse-response reparameterization, yielding a moderate data-aware adjustment that enhances low-response ranges while aiming to preserve relative spatial structure.

In our method, $F[u]$ denotes the average perceptual response associated with scalar-value bin $u$, rather than a histogram over the values of $T(x)$. The warp function $\phi$ is defined as a sampling warp on the colormap domain, so the adjusted colormap is written as $\tilde{\gamma}(t)=\gamma(\phi(t))$. 

\begin{figure}[t!]
    \centering
    \includegraphics[width=1.0\linewidth]{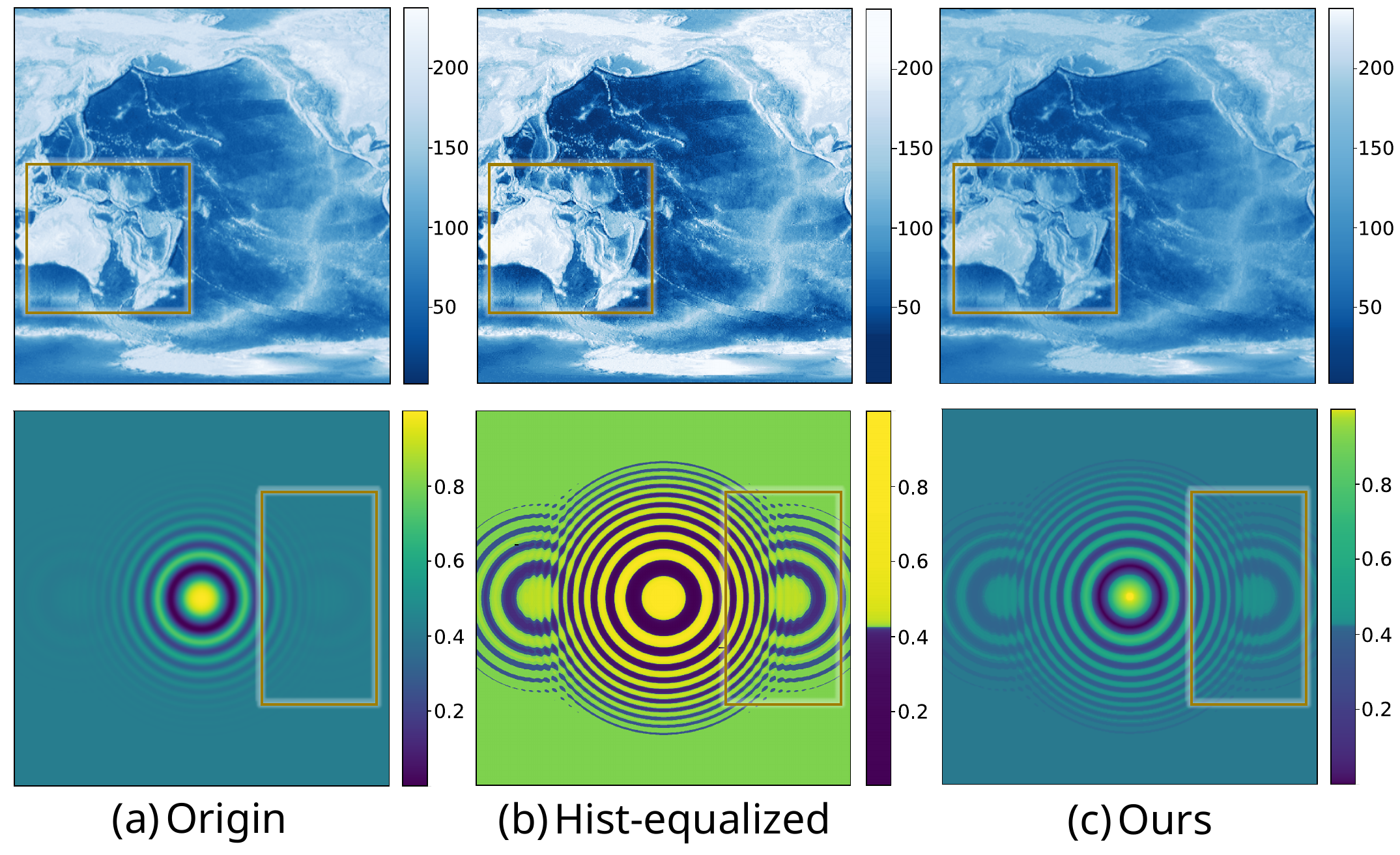}
    \caption{\textbf{Colormap enhancement results}. We test our colormap adjustment algorithm (c) and traditional histogram equalization (b) on \textit{viridis} and \textit{singlehue} with datasets \textit{ripples} (D5, top) and \textit{ocean salinity} (D6, bottom). With redistributed perceptual emphasis, the enhanced colormap preserves structural details while visually emphasizing regions that were previously under-resolved, compared to histogram equalization.} 
    \label{fig:exp-hist}
\end{figure}

As shown in \cref{fig:exp-hist}, we compare the original colormap (a), traditional histogram equalization (b), and our perceptual adjustment (c) on two colormaps (\textit{viridis}, \textit{singlehue}) and two datasets (\textit{ripples} D5, \textit{ocean salinity} D6). With settings $\alpha=0.5$ and 20 iterations, the adjusted colormaps preserve structural details while emphasizing regions that were previously under-resolved. Compared to traditional histogram equalization, which operates on data values without considering perceptual properties, our approach targets perceptual balance directly, guided by the $T(x)$ field. This case demonstrates that the data-aware measures can serve not only as evaluation tools but also as objective functions for colormap optimization, suggesting a promising direction for integrating data-aware colormap assessment into color design. Note that our method is not intended to maximize perceptual uniformity, as the problem is inherently a trade-off between uniformity and the enhancement of locally underrepresented regions. Further comparisons and discussions are provided in the supplementary material.

\begin{algorithm}[t!]
\caption{Perception Histogram-based Colormap Enhancement}
\begin{algorithmic}[1]
\Require 2D scalar field $f(x)$, region $\Sigma$, colormap $\gamma(t)$, perceptual metric $h$, warp strength $\alpha \in (0,1)$, number of bins $n$
\Ensure Enhanced colormap $\tilde\gamma(t)$

\State \textit{compute field} $T(x)$ \textit{using} $h$
\Comment{\cref{eq:tr}}

\State $F[u] \leftarrow$ \verb|weightedHist|$(f, T, n\text{-bins}, \Sigma)$
\Comment{bin-wise mean response}

\State $q(u) \leftarrow (1-\alpha)/n + \alpha / \sqrt{F[u]+\epsilon}$
\Comment{target allocation density}

\State $\phi(u) \leftarrow \sum_{v < u} q(v)$
\Comment{construct a monotone sampling warp}

\State \textit{normalize} $\phi$ $s.t.$ $\phi(0)=0$, $\phi(1)=1$

\State $\tilde\gamma(t) \leftarrow \gamma(\phi(t))$
\Comment{enhance colormap using the warp}

\State \Return $\tilde\gamma$
\end{algorithmic}
\label{alg:hist}
\end{algorithm}

\subsection{Interactive Exploration System}
\begin{figure}[t!]
    \centering
    \includegraphics[width=1.0\linewidth]{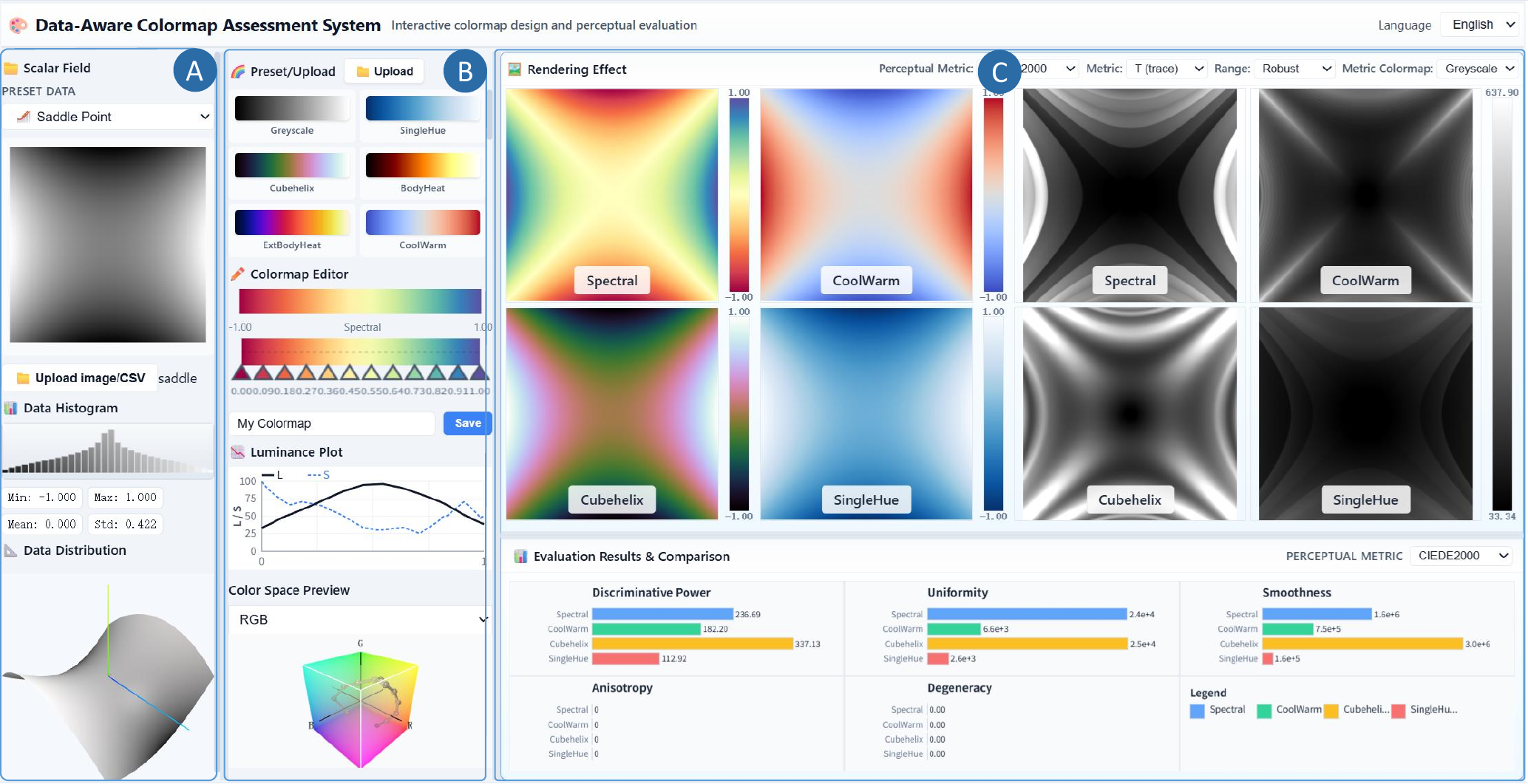}
    \caption{
\textbf{Interactive system for data-aware colormap assessment. }
The interface is organized into three components: 
(A) data specification with scalar field selection and distribution summary; 
(B) colormap selection and design with perceptual structure visualizations (e.g., luminance profiles and color space trajectories); 
(C) data-aware evaluation, where multiple colormaps are compared on the same dataset. 
The main view shows both color-encoded renderings and spatial maps derived from the perceptual metric, revealing how perceptual sensitivity varies across the data domain. 
These local patterns are aggregated into global measures (e.g., discriminative power and uniformity) for quantitative comparison.
}
    \label{fig:system}
\end{figure}

To support practical exploration of data-aware colormap assessment, we develop an interactive system that enables users to examine how colormap performance varies jointly with data, colormap design, and evaluation criteria (\cref{fig:system}). The system is organized around the core idea that colormap quality should be evaluated in the context of the data it encodes, and integrates three tightly coupled components: data specification (panel A), colormap design (panel B), and data-aware evaluation (panel C).

\vspace{0.3em}
\noindent\textbf{Data specification.} Users can select from predefined datasets or upload custom scalar fields. A histogram with summary statistics displays the value distribution at a glance.

\vspace{0.3em}
\noindent\textbf{Colormap design and analysis.} Users can choose from preset colormaps or construct custom ones via a control-point editor with real-time feedback. The system displays the colormap's trajectory in perceptual color spaces (e.g., RGB, HSV, CIELAB) alongside its luminance and saturation profiles, allowing users to inspect how perceptual properties vary along the colormap.

\vspace{0.3em}
\noindent\textbf{Data-aware Colormap Assessment.} The main view presents side-by-side visualizations of the same dataset under multiple colormaps. Beyond the rendered images, the system visualizes spatial maps of the perceptual measures, including $T(x)$, roughness $R(x)$, anisotropy $\Lambda(x)$, and degeneracy $G(x)$, revealing how perceptual sensitivity varies across the data. These local quantities are aggregated into global measures, including discriminative power, uniformity, smoothness, anisotropy, and degeneracy, computed under a user-selected color difference.

\vspace{0.3em}
\noindent\textbf{Interactions.} The system supports coordinated interactions across all components. Switching datasets updates all downstream views, exposing how colormap behaves under different data. Editing a colormap immediately updates both the spatial metric field maps and the global measures, forming a tight feedback loop between design and evaluation. Users can also define regions of interest for localized analysis.

\section{Discussions}
\paragraph{Scope and Assumptions.}
Our formulation is designed for continuous colormaps, where the colormap function $\gamma$ is differentiable and the Jacobian $J_\gamma$ is well-defined. Discrete colormaps, which assign distinct colors to intervals of the scalar range, are piecewise constant and thus not differentiable. The perceptual metric field in our formulation cannot be constructed in this setting. \emph{Extending the data-aware assessment to discrete colormaps would require a fundamentally different mathematical formulation and remains an open problem.} Furthermore, even for continuous colormaps, derivative computation contains discretization artifacts in numerical evaluation. We further discuss this in our supplementary material.

Among the continuous colormap measures formalized by Bujack~\etal~\cite{Bujack18}, our formulation derives data-aware counterparts of discriminative power, uniformity, and smoothness, but not order. The calculation of order requires that the perceptual speed remains non-negative along a designated perceptual channel, encoding a directional property: colors should progress monotonically as scalar values increase. The perceptual metric, however, is a quadratic form that measures the magnitude of perceptual change without retaining its sign. \emph{Incorporating order into the data-aware formulation would require preserving signed directional information through the chain rule, which goes beyond the standard metric formulation.}

Our formulation also rests on several additional assumptions. We assume a linear mapping between data values and colors, which is the standard convention in most visualization tools~\cite{Tominski08, Bujack18, Liu25}. Non-linear mappings, such as gamma correction or logarithmic scaling, are not explicitly modeled, but can be incorporated by introducing an additional transformation $\phi$ into the composite mapping, yielding $\Phi = \gamma \circ \phi \circ f$, with the corresponding Jacobian $J_\phi$ entering the derivative computation. \emph{Incorporating such transformations would broaden the formulation's applicability to visualization pipelines that employ histogram equalization, logarithmic scaling, or other non-linear mappings.} 

Following prior work~\cite{Pant10, Molchanov25}, we assume that the perceptual color space is locally Riemannian, allowing us to define the perceptual metric $h$ and compute derivatives. Recent work by Bujack~\etal~\cite{Bujack22} has shown that this assumption does not hold at large color differences. Since our formulation operates on local differences between neighboring points in the data domain, the Riemannian approximation remains appropriate for the current setting. \emph{Extending the formulation to non-Riemannian perceptual geometry would be a natural next step as more accurate models of large-scale color perception become available.} The current implementation focuses on 2D scalar fields, and while the mathematical formulation is dimension-independent, \emph{applying it to 3D volumetric or time-varying data would require addressing how to aggregate metric field information over higher-dimensional domains.}

\paragraph{Relationship to the Data-Independent formulation.}
Our data-aware formulations should complement, rather than replace, previous data-independent colormap evaluations~\etal~\cite{Bujack18}. As demonstrated in our experiments, the data-aware measures recover the classical data-independent definitions under uniform data, confirming that our formulation is a strict generalization of the existing framework. The two perspectives serve different but complementary purposes. Data-independent measures characterize the intrinsic perceptual properties of a colormap and provide a universal baseline for comparison that is valid regardless of what data the colormap will be applied to. Data-aware measures evaluate how those properties manifest when the colormap is applied to a specific dataset, capturing the joint effect of the colormap and the data on user perception. In practice, we envision the two being used together: data-independent measures for general-purpose colormap selection and quality assessment, and data-aware measures for evaluating and refining colormap choices in the context of a particular dataset or application. \emph{The integration of data-aware measures into colormap design, which we demonstrated with a preliminary colormap adjustment example (\cref{subsec:generalization}), deserves further development. The perceptual metric field provides a spatially resolved objective that could guide more sophisticated optimization strategies}.

\paragraph{Perceptual Validity.}
We validated the proposed measures against empirical results from two perceptual tasks reported by Reda~\etal~\cite{Reda18}: quantity estimation and gradient comparison. The results demonstrate that our data-aware measures align with observed perceptual performance, particularly under non-uniform data where data-independent measures are less predictive. However, this validation covers only two specific tasks, and colormap effectiveness in practice spans a broader range of activities, such as anomaly detection and value comparison at different spatial scales. \emph{Conducting more extensive user studies across a wider range of tasks and data types would strengthen the empirical grounding of the proposed measures.}
 
It is also worth noting that our measures are grounded in low-level perceptual color differences, which capture how distinguishable neighboring colors are to the human visual system. They do not account for higher-level cognitive factors that also influence colormap effectiveness, such as color name~\cite{Heer2012,Reda21,Lu2026}, color-concept associations~\cite{Mukherjee2022,Schoenlein2022,Schoenlein2023}, or color emotion~\cite{Bartram2017,Braun2026}. \emph{Investigating how these cognitive factors interact with the low-level color differences captured by our formulation could lead to richer, multi-level assessment models.}
 
\paragraph{Scalability and Practical Considerations.}
The computation of the perceptual metric requires evaluating the Jacobians of both the data field and the colormap at every point in the domain. For the 2D scalar fields considered in this work, this computation supports interactive exploration, as demonstrated by our system. For larger datasets, higher-dimensional fields, or real-time applications, computational cost may become a concern. \emph{Strategies such as multi-resolution evaluation, sampling-based approximation, or GPU-accelerated computation could be explored to address scalability.}
 
From a practical standpoint, the ability to evaluate measures within user-defined regions of interest is particularly valuable~\cite{Elmqvist11, Zeng22}. In many applications, users are not equally interested in all parts of the data domain. A meteorologist may focus on a storm region within a global pressure field; a radiologist may examine a specific tissue boundary in a medical scan. Region-based evaluation allows the data-aware measures to provide targeted guidance for colormap selection in these scenarios, rather than averaging over potentially irrelevant areas. \emph{Integrating such region-based evaluation with domain-specific task models could further improve the practical relevance of the assessment.}

\section{Conclusion}
We presented a data-aware formulation for continuous colormap assessment that extends established data-independent measures by incorporating the underlying data into the evaluation process. We modeled color-encoded visualization as a composite mapping from the data manifold to perceptual color space, and defined a data-aware perceptual metric that quantifies how local data variations translate into perceptual differences. From this metric field, we derived data-aware counterparts of discriminative power, uniformity, and smoothness, as well as additional diagnostic measures including perceptual anisotropy and degeneracy. We validated the measures through evaluations, including quantitative comparisons with the data-independent framework, consistency checks against empirical perceptual studies, and extensions to 2D colormaps and colormap optimization. We also developed an interactive system that supports exploration of colormap performance under varying data and evaluation criteria. Our results suggest that jointly considering the colormap and the data it encodes provides a more complete picture of colormap effectiveness, offering a principled basis for both evaluation and design in practical visualization settings.

\acknowledgments{
This study is financially supported by Laoshan National Laboratory (No.~LSKJ202601904) and a grant from the National Natural Science Foundation of China (No.~62372271).
}

%% if specified like this the section will be omitted in review mode

\bibliographystyle{abbrv-doi-hyperref}

\bibliography{ref}

% \appendix % You can use the `hideappendix` class option to skip everything after \appendix

% \section{Appendix}
% \section{About Appendices}
% If you receive compilation errors along the lines of \texttt{Package ifpdf Error: Name clash, \textbackslash ifpdf is already defined} then please add a new line \verb|\let\ifpdf\relax| right after the \verb|\documentclass[journal]{vgtc}| call.
% Note that your error is due to packages you use that define \verb|\ifpdf| which is obsolete (the result is that \verb|\ifpdf| is defined twice); these packages should be changed to use \verb|ifpdf| package instead.

% \subsection{\texttt{pdfendlink} error}
% Occasionally (for some \LaTeX\ distributions) this hyper-linked bib\TeX\ style may lead to \textbf{compilation errors} (\texttt{pdfendlink ended up in different nesting level ...}) if a reference entry is broken across two pages (due to a bug in \verb|hyperref|).
% In this case, make sure you have the latest version of the \verb|hyperref| package (i.e.\ update your \LaTeX\ installation/packages) or, alternatively, revert back to \verb|\bibliographystyle{abbrv-doi}| (at the expense of removing hyperlinks from the bibliography) and try \verb|\bibliographystyle{abbrv-doi-hyperref}| again after some more editing.

\end{document}